\documentclass[10pt,english]{article}
\usepackage{times}
\usepackage[T1]{fontenc}
\usepackage[latin1]{inputenc}
\usepackage{graphicx}
\usepackage{multicol}
\usepackage{babel}
\usepackage{geometry}
\usepackage{hyperref}
\makeatletter
\renewcommand\maketitle
{
  \begin{center}%
    {\LARGE \@title \par}%
    \vskip 1em%
    {\large
      \lineskip .75em%
      \begin{tabular}[t]{c}%
        \@author
      \end{tabular}\par}%
  \end{center}
  \vskip 1em%
  \setcounter{footnote}{0}%
  \global\let\maketitle\relax \global\let\@author\@empty
  \global\let\@title\@empty \global\let\title\relax
  \global\let\author\relax \global\let\and\relax
}
\makeatother
\begin{document}
%% Title
\title{\textbf{Studies of Short-Time Decoherence for\\
Evaluation of Quantum Computer Designs\\ }}
\author{{\bf Vladimir Privman \ and \ Denis Tolkunov}\\
\vphantom{$T^{T^{T^{T^a}}}$}Center for Quantum Device Technology, Clarkson University, \\Potsdam, New York 13699--5721, USA\\
\vphantom{$T^{T^{T^{T^a}}}$}E-mails:\ privman@clarkson.edu,\ tolkunov@clarkson.edu
} \maketitle
%% Main text: two columns, no page numbers
\thispagestyle{empty}
\setlength{\columnsep}{0.8cm}\begin{multicols}{2}
\section*{Abstract}

We review our recent results on short time approximations, with
emphasis on applications for which the system-environment
interactions involve a general  non-Hermitian system
operator $\Lambda$, and its conjugate, $\Lambda^\dagger$. We
evaluate the onset of decoherence at low temperatures in open quantum
systems. The developed approach is
complementary to Markovian approximations and appropriate for
evaluation of quantum computing schemes. Example of a spin system
coupled to a bosonic heat bath via $\Lambda \propto \sigma_-$ is
discussed. %%

\section{\label{sec:level1}Introduction}

The coupling of a quantum system to environmental degrees of
freedom induces decoherence, destroying quantum superposition and
reducing pure states to mixed states. Understanding decoherence
is important for quantum control and computing and,
generally, for obtaining a description to the evolution
of the system's reduced density matrix with the environmental
modes traced over. Since this can not be done exactly in most cases, 
different approximation techniques which are valid for
different time scales were developed. For short times, appropriate
for quantum computing gate functions and, generally, for
controlled quantum dynamics, approximation schemes for the density
matrix have been suggested recently
\vphantom{\cite{Tolk,Loss,Onset1,Onset2}}\hbox{[1--4]}. The present
survey introduces a new \cite{Tolk} rather general short-time approximation which
applies for models with system-bath interactions involving a
general system operator. It thus extends the previously known
approach \vphantom{\cite{Onset1,Onset2}}[3,4] developed for
couplings involving a single Hermitian system operator.

$\quad$We assume that the Hamiltonian of the open quantum system is
\begin{equation}
H=H_S+H_B+H_I,  \label{eq:HHH}
\end{equation}
where $H_S$ describes the system which is coupled to a fluctuating
dynamical reservoir (the bath). 
Typically, the bath is modeled by the harmonic modes, as reviewed in
\cite{Legg},
\begin{equation}
H_B=\sum\limits_k \omega _k^{\vphantom{\dagger}} b_k^{\dagger
}b_k^{\vphantom{\dagger}}.
\end{equation}
Here $b_k$ are the annihilation operators of
the bath modes, and we use the convention $\hbar=1 $.
We assume that the interaction with the bath involves the system
operator $\Lambda$ that couples linearly to the bath
modes, as reviewed in \cite{Grab}, 
\begin{equation}
H_I = \Lambda \sum\limits_k g_k^{\vphantom{\dagger}} b_k^{\dagger
} + \Lambda ^{\dagger } \sum\limits_k g_k^{*}
b_k^{\vphantom{\dagger}},  \label{eq:one}
\end{equation}
with the interaction constants $g_k$.

$\quad$Let $R(t)$ denote the overall density matrix. It is 
assumed \vphantom{\cite{Onset1,VKam2}}[3,7] that at time $t=0$ the system and
bath are not entangled, and the bath modes are thermalized ($ \beta\equiv 1 / kT $):
\begin{equation}\label{RR}
R\left( 0\right) =\rho \left( 0\right) \prod\limits_k\theta _k,
\end{equation}
\begin{equation}\label{theta}
\theta _k \equiv ( 1-e^{-\beta \omega _k} ) \, e^{-\beta \omega
_k^{\vphantom{\dagger}}b_k^{\dagger }b_k^{\vphantom{\dagger}}}.
\end{equation}

$\quad$We point out that while the quantum system $S$, described by the
reduced density matrix $\rho(t)=\mathrm{Tr}_B R\left( t\right)$, 
is small, typically two-state
(qubit) or several-qubit, the bath has many degrees of freedom.
The combined effect of the bath modes on the system can be large
even if each of them is influenced little by the system. This has
been the basis for the arguments for the harmonic approximation
for the bath modes and the linearity of
the interaction, as well as for the Markovian approximations
\vphantom{\cite{Grab,VKam2}}[6,7] that assume that the bath modes are
``reset'' to the thermal state by the ``rest of the universe'' on
time scales shorter than any dynamical time of the system
interacting with the bath.

$\quad$The frequencies of the oscillators of the bath are usually assumed
to be distributed from zero to some cutoff value $\omega_c$. The
bath modes with the frequencies close to the energy gaps of the
system, $\Delta E_{ij}=E_i-E_j$, contribute to the ``resonant''
thermalization and decoherence processes. Within the Markovian
schemes, the diagonal elements of the reduced density matrix of
the system,
approach the thermal values $\propto e^{-E_i/kT} $ for large times
exponentially, on time scale $T_1$. The off-diagonal elements
vanish, which represents decoherence, on time scale $T_2$, which,
for resonant processes, is given by $T_2 \simeq 2T_1$. However,
generally decoherence is expected to be faster than thermalization
because, in addition to resonant processes, it can involve virtual
processes that do not conserve energy. It has been argued that
this additional ``pure'' decoherence is dominated by the bath
modes with near-zero frequencies
\vphantom{\cite{Onset1,Grab,VKam2}}[3,6,7]. At low temperatures, this
``pure decoherence'' is expected \cite{12} to make $T_2 \ll T_1$.

$\quad$Since the resetting of these low-frequency modes to the thermal
state occurs on time scales $\hbar/kT=\beta$, the Markovian approach
cannot be used at low temperatures
\vphantom{\cite{Onset1,VKam2}}[3,7]. For quantum
computing in semiconductor-heterostructure
architectures
\vphantom{\cite{12,16,17,18,19,20,21,22}}[8--15],
temperatures as low as few $10${\,}mK are needed. This brings the
thermal time scale to $\beta \sim 10^{-9}\,$sec, which is close to
the single-qubit control times $10^{-11} \hbox{-} 10^{-7}\,$sec. 
Alternatives to the Markovian approximation
have been suggested
\vphantom{\cite{Tolk,Loss,Onset1,Onset2,Ford,Romero1,Romero2,OConnell,Shnirman,Scnon}}[1--4,16--21].

\section{Short-Time Approximation}

In applications in
quantum computing, calculations with only a single qubit or few
qubits are necessary for evaluation of the local ``noise,'' to use
the criteria for quantum error correction
\vphantom{\cite{Shor,29,30,31,32,Presk}}[22--27].
For example, the system Hamiltonian is frequently taken
proportional to the Pauli matrix $\sigma_z$. The interaction
operator $\Lambda$ can be proportional to $\sigma_x$, which is
Hermitian. Such cases are covered by the short-time approximation
developed earlier \vphantom{\cite{Onset1,Onset2}}[3,4]. However, one can
also consider models with $\Lambda \propto \sigma_-$. Similarly,
models with non-Hermitian $\Lambda$ are encountered in Quantum
Optics \cite{Lois}. In this section, we develop our short time
approximation scheme. Results for a spin-boson type model are
given in the next section.

$\quad$We derive a general expression for the time evolution operator of
the system (\ref{eq:HHH})--(\ref{eq:one}) within the short time
approximation. 
The overall density matrix, assuming time-independent Hamiltonian
over the quantum-computation gate function time intervals
\vphantom{\cite{12}-\cite{22}}[8--15], evolves according to
\begin{equation}
R\left( t\right) =U(t)R\left( 0\right) [U(t)]^{\dagger },
\end{equation}
\begin{equation}
U(t) \equiv e^{-i\left( H_S+H_B+H_I\right) t} . \label{eq:evolut}
\end{equation}
The general idea of our approach is to break the
exponential operator in (\ref{eq:evolut}) into products of simpler
exponentials. This involves an approximation, but allows us to
replace system operators by their eigenvalues, when spectral
representations are used, and then calculate the trace of $R(t)$
over the bath modes, obtaining explicit expressions for the
elements of the reduced density matrix of the system. For
Hermitian coupling operators, $\Lambda^{\dagger}=\Lambda$, our
approach reduces to known results \vphantom{\cite{Onset1,Onset2}}[3,4].

$\quad$In order to define ``short time,'' we consider dimensionless
combinations involving the time variable $t$. There are several
time scales in the problem. These include the inverse of the
cutoff frequency of the bath modes, $1 / {\omega _c}$, the thermal
time $\beta= 1 / {kT}$, and the internal characteristic times of
the system $1 / {\Delta E_{ij}}$. Also, there are time scales
associated with the system-bath interaction-generated
thermalization and decoherence, $T_{1,2}$. The shortest time scale
at low temperatures (when $\beta$ is large) is typically $1 /
{\omega _c}$. The most straightforward expansion in $t$ yields a
series in powers of ${\omega _c} t $. The aim of developing more
sophisticated short-time approximations
\vphantom{\cite{Tolk,Onset1,Onset2}}[1,3,4] has been to preserve unitarity and obtain
expressions approximately valid up to intermediate times, of order
of the system and interaction-generated time scales. The applicability 
for intermediate times  can only be argued for heuristically in 
most cases, and checked by model calculations.

$\quad$We split the exponential evolution operator into terms that do not
have any noncommuting system operators in them. This requires an
approximation. For short times, we start by using the
factorization 
\begin{eqnarray}
&&e^{-i\left( H_S+H_B+H_I\right) t+O\left( t^3\right) }  \nonumber
\label{HB} \\
&=&e^{-\frac i2H_St}e^{-i\left( H_I+H_B\right) t}e^{-\frac
i2H_St},
\end{eqnarray}
where we have neglected terms of the third and higher orders in
$t$, in the exponent. The middle exponential in (\ref{HB}),
\begin{equation}
\Xi \equiv e^{-i\left( H_B+H_I\right) t}=e^{-i\left( H_B+\Lambda
G^{\dagger }+\Lambda ^{\dagger }G\right) t},
\end{equation}
where
\begin{equation}
G\equiv \sum\limits_kg_k^{*}b_k,
\end{equation}
still involves noncommuting terms as long as $\Lambda$ is
non-Hermitian. In terms of the Hermitian operators
\begin{eqnarray}
&&L\equiv {\frac 12}\left( \Lambda +\Lambda ^{\dagger }\right) ,  \label{lam1} \\
&&M\equiv {\frac i2}\left( \Lambda -\Lambda ^{\dagger }\right) ,
\label{lam2}
\end{eqnarray}
we have
\begin{equation}
\Lambda G^{\dagger }+\Lambda ^{\dagger }G=L\left( G+G^{\dagger
}\right) +iM\left( G-G^{\dagger }\right) .
\end{equation}
We then carry out two additional short-time factorizations within
the same quadratic-in-$t$ (in the exponent) order of
approximation,
\begin{eqnarray}\label{eq:decomp3}
\Xi &= & e^{\frac 12\left[ M\left( G-G^{\dagger }\right)
-iH_B\right] t}e^{\frac i2H_Bt}\\\nonumber &\times&e^{-i\left[
H_B+L\left( G+G^{\dagger }\right) \right] t}e^{\frac
i2H_Bt}e^{\frac 12\left[ M\left( G-G^{\dagger }\right)
-iH_B\right] t}.
\end{eqnarray}
This factorization is chosen in such a way that $\Xi$ remains
unitary, and for $M=0$ or $L=0$ the expression is identical to
that used for the Hermitian case \vphantom{\cite{Onset1,Onset2}}[3,4]. The
evolution operator then takes the form
\begin{equation}
U=e^{-\frac i2H_St}\,\Xi \,e^{-\frac i2H_St},  \label{uu}
\end{equation}
with $\Xi$ from (\ref{eq:decomp3}), which is an approximation in
terms of a product of several unitary operators.

$\quad$It has been recognized \vphantom{\cite{Tolk,Onset1,Onset2}}[1,3,4] that
approximations of this sort, which are
not perturbative in powers of $H_I$, are superior to the straightforward
expansion in powers of $\omega_c t$.
Specifically, in (\ref{HB}), we notice that $H_S$ is factored out
in such a way that $H_B$, which commutes with $H_S$, drops out
of many commutators that enter the higher-order correction
terms. This suggests that a redefinition of the energies of the
modes of $H_B$ should have only a limited effect on the
corrections and serves as a heuristic argument for the
approximation being valid beyond the shortest time scale
$1/\omega_c$, up to intermediate time scales.

$\quad$Our goal is to approximate the reduced density matrix of the
system. We consider its energy-basis matrix elements,
\begin{equation}
\rho _{mn}\left( t\right) =\mathrm{Tr}_B\left\langle m\right|
UR\left( 0\right) U^{\dagger }\left| n\right\rangle ,  \label{rho}
\end{equation}
where
\begin{equation}
H_S\left| n\right\rangle =E_n\left| n\right\rangle .  \label{eig}
\end{equation}
We next use the factorization (\ref{HB}), (\ref{eq:decomp3}) to
systematically replace system operators by c-numbers, by inserting
decompositions of the unit operator in the bases defined by $H_S$,
$L$, and $M$. First, we collect the expressions
(\ref{RR}), (\ref{eq:decomp3}), (\ref{uu}), (\ref{eig}), and use two
energy-basis decompositions of unity to get
\begin{eqnarray}\label{NS}
\rho _{mn}\left( t\right) &=&\sum\limits_{p\,q}e^{\frac i2\left(
E_n+E_q-E_m-E_p\right) t}\rho _{pq}\left( 0\right)  \nonumber   \\
&\times& \mathrm{Tr}_B\left[ \left\langle m\right| \Xi \left|
p\right\rangle \right. \prod\limits_k\theta _k\left. \left\langle
q\right| \Xi ^{\dagger }\left| n\right\rangle \right] .
\end{eqnarray}

$\quad$We define the eigenstates of $L$ and $M$,
\begin{equation}
L\left| \lambda \right\rangle =\lambda \left| \lambda
\right\rangle ,
\label{eigen1} \\
M\left| \mu \right\rangle =\mu \left| \mu \right\rangle .
\label{eigen2}
\end{equation}
The operators $\Xi $ and $\Xi ^{\dagger }$ introduce exponentials
in (\ref{NS}) that contain $L$ or $M$ in the power. By
appropriately inserting $\sum\limits_\lambda \left| \lambda
\right\rangle \left\langle \lambda \right| $ or $\sum\limits_\mu
\left| \mu \right\rangle \left\langle \mu \right| $ between these
exponentials, we can convert all the remaining system operators to
c-numbers.

$\quad$Now the trace in (\ref{NS}) can be evaluated, by using operator
identities for bosonic operators \cite{Lois} and the coherent-states
technique. We obtain our final result for the density matrix
evolution \cite{Tolk},
\begin{eqnarray}\label{eq:EV} \nonumber
\rho _{mn}\left( t\right)  &=&\sum\limits_{p,q}\sum\limits_{\mu
_j\lambda _j}e^{\frac i2\left( E_n+E_q-E_m-E_p\right)
t-\mathcal{P}}\rho _{pq}\left( 0\right)\\\nonumber &\times
&\left\langle m\right| \left. \mu _1\right\rangle
  \left\langle \mu _1\right| \left. \lambda
_1\right\rangle \left\langle \lambda _1\right. \left| \mu
_2\right\rangle \left\langle \mu _2\right| \left. p\right\rangle\\
&\times & \left\langle q\right| \left. \mu _3\right\rangle
\left\langle \mu _3\right| \left. \lambda _2\right\rangle
\left\langle \lambda _2\right. \left| \mu _4\right\rangle
\left\langle \mu _4\right| \left. n\right\rangle,
\end{eqnarray}
where the first sum over $p$ and $q$ is over the energy
eigenstates of the system; the second sum is over $\lambda_1
,\lambda_2$ and $\mu_1 , \ldots , \mu_4$, which label the
eigenstates of the operators $L$ and $M$, respectively. The power
in the exponential is
\begin{eqnarray}
\mathcal{P}&=&B^2\left( t\right) \left( \lambda _{-}^2+\mu
_{-}^{\prime }\mu _{-}^{\prime \prime }\right) +B^2\left(
t/2\right) \left( \mu _{-}^{\prime \prime }-\mu _{-}^{\prime
}\right) ^2\nonumber\\&-&F\left( t\right) \left( \mu _{-}^{\prime
\prime }-\mu _{-}^{\prime }\right) \lambda _{-} - iC\left(
t\right) \lambda _{-}\lambda _{+}-iC\left(
t/2\right)\nonumber\\&\times& \left( \mu _{-}^{\prime }\mu
_{+}^{\prime }\right.\nonumber+\left.\mu _{-}^{\prime \prime }\mu
_{+}^{\prime \prime }\right) +iS\left( t\right) \left( \lambda
_{-}\mu _{+}^{\prime \prime }-\lambda _{+}\mu _{-}^{\prime
}\right)\nonumber\\ &-&iC_1\left( t\right) \mu _{-}^{\prime }\mu
_{+}^{\prime \prime }.
\end{eqnarray}
Here we introduced the variables
\begin{eqnarray}
\mu _{\pm }^{\prime } &=&\mu _1\pm \mu _4 , \\
\mu _{\pm }^{\prime \prime } &=&\mu _2\pm \mu _3,
\end{eqnarray}
and
\begin{equation}
\lambda _{\pm }=\lambda _1\pm \lambda _2.
\end{equation}
and the spectral sums over the bath
modes,
\begin{eqnarray}\label{spec1}
B^2\left( t\right) =2\sum\limits_k\frac{\left| g_k\right|
^2}{\omega _k^2} \sin ^2\frac{\omega _kt}2\coth \frac{\beta \omega
_k}2,
\end{eqnarray}
\begin{eqnarray}\label{spec2}
C\left( t\right) =\sum\limits_k\frac{\left| g_k\right| ^2}{\omega
_k^2} \left( \omega _kt-\sin \omega _kt\right) ;
\end{eqnarray}
these functions are well known \vphantom{\cite{35,27}}[29,30]. The result
also involves the new spectral functions
\begin{eqnarray}\label{spec3}
S\left( t\right) =-2\sum\limits_k\frac{\left| g_k\right|
^2}{\omega _k^2}\sin ^2\frac{\omega _kt}2,
\end{eqnarray}
\begin{eqnarray}\label{spec4}
F\left( t\right) =4\sum\limits_k\frac{\left| g_k\right| ^2}{\omega
_k^2}\sin ^2\frac{\omega _kt}4\sin \frac{\omega _kt}2\coth
\frac{\beta \omega _k}2.
\end{eqnarray}
Furthermore, we defined
\begin{equation}\label{C1}
C_1\left( t\right) =2C\left( t/2\right) -C\left( t\right) .
\end{equation}

\section{Discussion and Application}

In most applications evaluation of decoherence will
require short-time expressions for the reduced density matrix of a
single qubit. Few- and multi-qubit systems will have to be treated
by utilizing additive quantities
\vphantom{\cite{norm,dd,addnorm}}\hbox{[31--33]}, accounting for quantum error
correction (requiring measurement), etc. For a two-state
system---a qubit---the summation in (\ref{eq:EV}) involves
$2^8=64$ terms, each a product of several factors calculation of
which is straightforward. We
carry out the calculation for an illustrative example.

$\quad$We consider the model \cite{Swain} defined by
\begin{equation}
H=\mathcal{A}{\,}\sigma _z+\sum\limits_k\omega
_k^{\vphantom{\dagger} }b_k^{\dagger
}b_k^{\vphantom{\dagger}}+\sum\limits_k\left(
g_k^{\vphantom{\dagger}}\sigma _{-}b_k^{\dagger }+g_k^{*}\sigma
_{+}b_k^{\vphantom{\dagger}}\right) ,  \label{RWA}
\end{equation}
where $\mathcal{A} \geq 0$ is a constant, $\sigma _{\pm}={1\over
2} (\sigma_x \pm i\sigma_y)$ and $\sigma _z$ are the Pauli
matrices, $b_k^{\dagger }$ and $b_k^{\vphantom{\dagger}}$ are the
bosonic creation and annihilation operators, and $g_k$ are the
coupling constants. Physically this model may describe, for
example, a qubit interacting with a bath of phonons, or a
two-level molecule in an electromagnetic field. In the latter
case, this is a variant of the multi-mode Jaynes-Cummings model
\vphantom{\cite{Lois,JC}}[28,35]. Certain spectral properties of this model,
the field-theoretic counterpart of which is known as the Lee field
theory, are known analytically, e.g., \cite{Pf}. However, the
trace over the bosonic modes, to obtain the reduced density matrix
for the spin, has not been calculated exactly.

$\quad$For the model (\ref{RWA}) we have $\Lambda =\sigma _{-}$ and
$\Lambda ^{\dagger }=\sigma _{+}$, so that $L=\sigma _x/2$ and
$M=\sigma _y/2.$ We have $\left| \lambda _{1,2 }\right\rangle
=\left( \left| \uparrow \right\rangle \pm \left| \downarrow
\right\rangle \right) /\sqrt{2}$, with the eigenvalues $ \lambda _{1,2
}=\pm 1/2$, and $\left| \mu _{1,2 }\right\rangle =\left(\left|
\uparrow \right\rangle \pm i\left| \downarrow \right\rangle
\right) / \sqrt{2}$, with the eigenvalues $\mu _{1,2 }=\pm 1/2$. For
the initial state, let us assume that the spin at $t=0$ is in the
excited state $\left| \uparrow \right\rangle \left\langle \uparrow
\right| $, so that the initial density matrix has the form
\begin{equation}
\rho \left( 0\right) =\left(
\begin{array}{cc}
1 & 0 \\
0 & 0
\end{array}
\right) .
\end{equation}
Our calculation yielded the following results for the
density matrix elements: $\, \rho_{12}(t)=0$ and
\begin{eqnarray}\label{Density}
4\rho _{11}\left( t\right)&=&2+\,e^{-2B^2\left( t\right) }+
e^{-4B^2\left( \frac t2\right) }\cosh \left( 2\,F\right)\nonumber\\
&+&2e^{-2B^2\left( \frac t2\right) }\sinh \left( B_1\right) \cos
\left( S\right)+2e^{-B^2\left( \frac t2\right) }\nonumber \\
&\times&\cos \left( C_1\right) \sin \left( S\right) \nonumber
+ie^{-B^2\left( t\right) -B^2\left( \frac t2\right)
}\nonumber\left[ \,e^{iC_1}\right.\\&\times&\left.\sinh \left(
-iS+F\right) +\,e^{-iC_1}\sinh \left( -iS-F\right) \right],
\nonumber
\end{eqnarray}
where $C_1$ was defined in (\ref{C1}) and
\begin{equation}
B_1(t)=2B^2\left( t/2\right) -B^2\left( t \right) .
\end{equation}
Where not explicitly shown, the argument of all the spectral
functions entering (\ref{Density}) is $t$.

$\quad$In order to obtain irreversible behavior and evaluate a measure of
decoherence, we consider the continuum limit of infinitely many
bath modes. We introduce the density of the bosonic bath states
$\cal{D}\left(\omega\right)$, incorporating a large-frequency
cutoff $\omega_c$, and replace the summations in
\hbox{(\ref{spec1})--(\ref{spec4})} by integrations over $\omega$
\vphantom{\cite{Legg,35,VKam,52}}[5,29,37,38]. For instance,
(\ref{spec1}) takes the form,
\begin{eqnarray}
\ B^2\left( t\right) =\int\limits_0^\infty d\omega
\frac{{\cal{D}}\left( \omega \right) |g( \omega )|^2 }{\omega
^2}\sin ^2\frac{\omega t}2\coth \frac{\beta \omega }2   .
\end{eqnarray}
We will use the standard Ohmic-dissipation \cite{Legg} expression,
with an exponential cutoff, for our illustrative calculation,
\begin{eqnarray}
{\cal{D}}\left( \omega \right) |g( \omega )|^2 =\Omega \, \omega
\, e^{-\omega /\omega_c},
\end{eqnarray}
where $\Omega $ is a constant.

$\quad$Our results for the density matrix
elements depend on the dimensionless variable $\omega_c t$, as
well as on the dimensionless parameters $\Omega$ and $\omega_c
\beta$ ($= \hbar \omega_c / kT$, where we remind the reader that
$\hbar$, set to 1, must be restored in the final results).
Interestingly, the results do not depend explicitly on the energy
gap parameter $\mathcal{A}$, see (\ref{RWA}). This illustrates the
point that short-time approximations do not capture the
``resonant'' relaxation processes, but rather only account for
``virtual'' relaxation/decoherence processes dominated by the
low-frequency bath modes. However, the short-time approximations
of the type considered here are meaningful only for systems with
well-defined separation of the resonant vs.\ virtual decoherence
processes, i.e., for $\hbar / \mathcal{A} \gg 1/\omega_c$. For
such systems, $\hbar / \mathcal{A} = 1 / \mathcal{A}$ defines one
of the ``intermediate'' time scales beyond which the approximation
cannot be trusted.

$\quad$As an example, we calculated a measure of deviation of a qubit
from a pure state in terms of the ``linear entropy''
\vphantom{\cite{norm,addnorm,24}}[31,33,39],
\begin{equation}
s(t)=1-\mathrm{Tr}\left[\rho ^2\left( t\right)\right].
\end{equation}
Figure\ 1 schematically illustrates the behavior of $s(t)$ for
different $\Omega$ values, for the case $\omega_c^{-1} <<  \beta
$. The values of $s(t)$ increase from zero, corresponding to a
pure state, to $1/2$, corresponding to a completely mixed state,
with superimposed oscillations. For Ohmic dissipation, three time
regimes can be identified \cite{27}. The shortest time scale is
set by $t<O\left(1/{\omega _c}\right) $. The quantum-fluctuation
dominated regime corresponds to $O\left(1/{\omega _c}\right)
<t<O\left( 1/{kT}\right) $. The thermal-fluctuation dominated
regime is $t>O\left( 1/{kT}\right) $. Our short time approximation
yields reasonable results in the first two regimes. For $t>O\left(
1/{kT}\right) $ it cannot correctly reproduce the process of
thermalization. Instead, it predicts approach to the maximally
mixed state.

$\quad$Figure\ 2 corresponds to the parameter values typical for low
temperatures and appropriate for quantum computing applications,
$\omega_c \beta=10^3$, with $\Omega=1.5\cdot 10^{-7}$ chosen to
represent weak enough coupling to the bath to have the decoherence
measure reach the threshold for fault-tolerance, of order
$10^{-6}$, for ``gate'' times well exceeding $1/ \omega_c$, here
for $ \omega_c t $ over 10. The leading-order quadratic expansion
in powers of the time variable $t$, the validity of which is limited to $t<O\left(1/{\omega _c}\right) $, is also shown for comparison.

\vphantom{A}

\begin{center}
  \includegraphics[width=6.5cm, keepaspectratio]{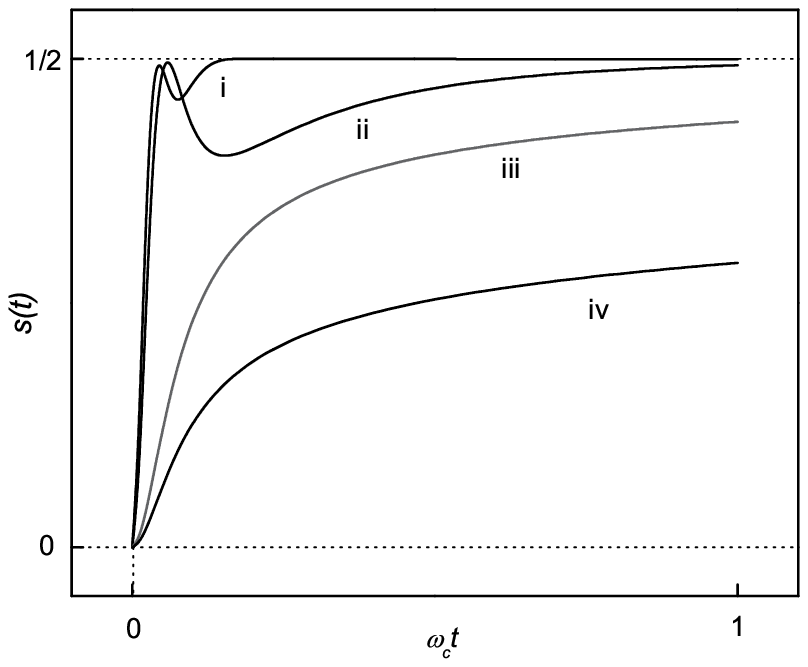}
\end{center}
\begin{center}\label{graph2}
{\bf Figure 1. }Schematic behavior of $s\left(t\right)$ for different
values of  $\Omega$, decreasing from $\sf{i}$ to $\sf{iv}$.
\end{center}

\vphantom{A}

\begin{center}
  \includegraphics[width=6.5cm, keepaspectratio]{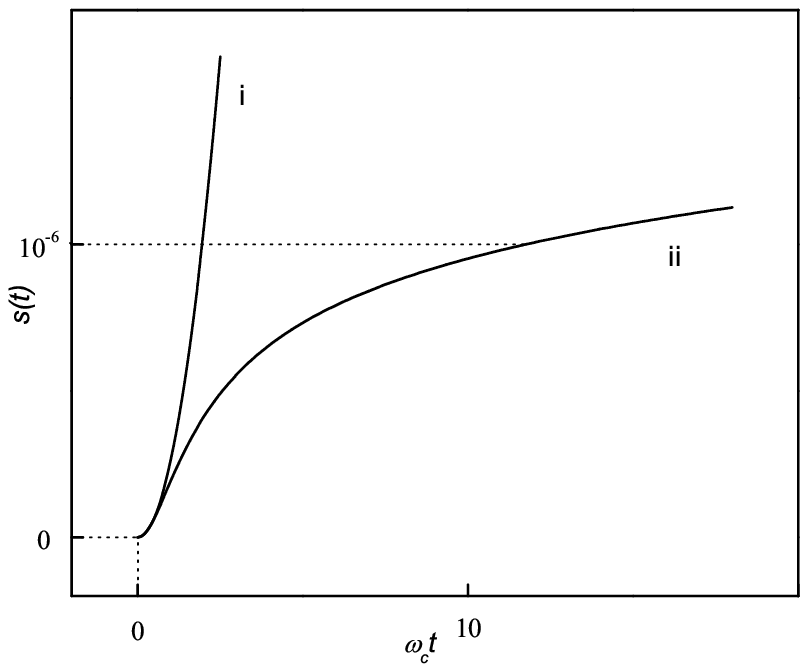}
\end{center}
\begin{center}\label{graph2}
{\bf Figure 2. }Comparison between the $O(t^2)$ expansion, $\sf{i}$,
and the short-time approximation, $\sf{ii}$.
\end{center}

\vphantom{A}

\vphantom{A}

\section*{Acknowledgement}
This research was supported by NSA and
ARDA under ARO contract DAAD-19-02-1-0035, and by NSF under grant DMR-0121146.\newline 
\hphantom{AAAAAAAAAAAAAAAAAAAAAAAAAAAAAAAAAAAAAAAAAA}\eject

%% Bibliography
{\frenchspacing
\bibliographystyle{icsict2004}
\bibliography{References} }
%% End
\end{multicols}
\end{document}